\documentclass[a4paper,pdftex,floatfix,groupedaddress,nofootinbib,pra,showkeys,showpacs,twocolumn]{revtex4}

\usepackage{graphicx}
\usepackage{hypernat}
\usepackage[utf8]{inputenc}
\usepackage[hyperfootnotes=false]{hyperref}

\pdfoutput=1

\hypersetup{colorlinks,citecolor=black,urlcolor=blue}

\newlength{\figwidth}
\newcommand{\degree}{\ensuremath{^\circ}}%
\newcommand{\etal}{et al.}%
\newcommand{\hnooo}{\ensuremath{\text{HNO}_3}}%
\newcommand{\ie}{i.\,e.}%
\newcommand{\state}{\ensuremath{{}^2\Pi_{3/2},v=0,J=3/2}}%
\newcommand{\lfsstate}{\ensuremath{\state,f}}%
\newcommand{\hfsstate}{\ensuremath{\state,e}}%

\begin{document}

\title{Stark deceleration of OH radicals \\ in low-field-seeking and high-field-seeking quantum states}%
\author{Kirstin Wohlfart}%
\author{Frank Filsinger}%
\author{Fabian Grätz}%
\author{Jochen Küpper}%
\email[Author to whom correspondence should be addressed; electronic mail:
]{jochen@fhi-berlin.mpg.de}%
\author{Gerard Meijer}%
\affiliation{Fritz-Haber-Institut der Max-Planck-Gesellschaft, Faradayweg 4--6, 14195 Berlin,
   Germany}%
\date{\today}%
\begin{abstract}\noindent%
   The Stark deceleration of OH radicals in both low-field-seeking and high-field-seeking levels of
   the rovibronic \state{} ground state is demonstrated using a single experimental setup. Applying
   alternating-gradient focusing, OH radicals in their low-field-seeking \lfsstate{} state have been
   decelerated from 345~m/s to 239~m/s, removing 50~\% of the kinetic energy using only 27
   deceleration stages. The alternating-gradient decelerator allows to independently control
   longitudinal and transverse manipulation of the molecules. Optimized high-voltage switching
   sequences for the alternating-gradient deceleration are applied, in order to adjust the dynamic
   focusing strength in every deceleration stage to the changing velocity over the deceleration
   process. In addition we have also decelerated OH radicals in their high-field-seeking \hfsstate{}
   state from 355~m/s to 316~m/s. For the states involved, a real crossing of hyperfine levels
   occurs at 640~V/cm, which is examined by varying a bias voltage applied to the electrodes.
\end{abstract}
\maketitle%

\section{Introduction}
\label{sec:introduction}

Since the first demonstration of Stark deceleration~\cite{Bethlem:PRL83:1558} tremendous advances
have been made in the deceleration and trapping of molecules in low-field-seeking (lfs)
states~\cite{Bethlem:Nature406:491, Meerakker:PRL94:023004}. Meanwhile, several groups have
successfully implemented Stark decelerators and applied them for the deceleration of different
molecules, namely metastable CO \cite{Bethlem:PRL83:1558}, different isotopologues of NH$_3$
\cite{Bethlem:Nature406:491, Bethlem:PRA65:053416} and OH \cite{Meerakker:PRL94:023004,
   Bochinski:PRL91:243001}, NH \cite{Meerakker:JPB39:S1077}, SO$_2$ \cite{Jung:PRA74:040701}, and
H$_2$CO \cite{Hudson:PRA73:063404}. In several other laboratories Stark decelerators are planned or
currently under construction. These studies are limited to molecules in low-field-seeking (lfs)
states. However, large molecules exhibit practically only high-field-seeking states, and the
absolute ground state of any molecule is high-field seeking (hfs). Therefore, it is desirable to
extend Stark deceleration to hfs states. This can be achieved by using dynamic, alternating-gradient
(AG) focusing schemes in the deceleration process~\cite{Auerbach:JCP45:2160, Bethlem:PRL88:133003,
   Bethlem:JPB39:R263}. AG deceleration has successfully been applied in the deceleration of
metastable CO~\cite{Bethlem:PRL88:133003, Bethlem:JPB39:R263}, the heavy diatomic molecule
YbF~\cite{Tarbutt:PRL92:173002}, and different rotational states of the large polyatomic molecule
benzonitrile~\cite{Wohlfart:PRA77:031404}.

Using AG focusing the whole deceleration process is analogous to the operation of a
LINAC~\cite{Lee:AccPhys:2004}. Although the forces on the neutral molecules are typically eight
orders of magnitude weaker than those applied in charged particle accelerators, the Stark
deceleration process can be used to slow neutral molecules in practice (\emph{vide supra}). In
contrast to the LINAC, the focusing process for neutral molecules is quantum-state specific. This
has been exploited, for example, for the separation of individual conformers of large
molecules~\cite{Filsinger:PRL100:133003}. Analogous to the LINAC, the phase stability of the Stark
deceleration process~\cite{Bethlem:PRL84:5744, Bethlem:PRA65:053416} ensures that the velocity of a
selected part of the beam can be varied without loss, \ie, molecules within a certain initial volume
in phase space are kept together throughout the deceleration process, independent of the length of
the decelerator. The concept of phase stability also applies to optical analogs of the Stark
decelerator \cite{Dong:PRA69:013409, Fulton:NatPhys2:465, Fulton:JPB39:S1097}. For the electric
field deceleration of atoms and molecules in Rydberg states \cite{Vliegen:JPB38:1623}, schemes that
employ phase stability have also been brought forward~\cite{Vanhaecke:JPB38:S409}, and decelerated
Rydberg hydrogen atoms have been trapped using electrostatic fields~\cite{Hogan:PRL100:043001}.
Furthermore, the magnetic analog of the Stark decelerator, the multistage Zeeman
decelerator~\cite{Vanhaecke:PRA75:031402}, has been used to decelerate oxygen
molecules~\cite{Narevicius:PRA77:051401}.

The first description and experimental demonstration of phase stability in a Stark decelerator was
given by Bethlem \etal\ using a beam of metastable CO molecules~\cite{Bethlem:PRL84:5744}. This
initial model for phase stability describes the longitudinal motion of molecules in a Stark
decelerator and predicts its longitudinal acceptance. Recently, an extended model for longitudinal
phase stability, including higher-order terms in the analysis, was
presented~\cite{Meerakker:PRA71:053409}. This model predicts a variety of additional phase-stable
regions, referred to as resonances, whose existence have been experimentally
verified~\cite{Meerakker:PRA71:053409}. The longitudinal motion has also been accurately described
by an analytical wave model~\cite{Gubbels:PRA73:063406}. In these models for longitudinal phase
stability, the motion of molecules through a Stark decelerator is treated one-dimensionally, \ie,
the trajectories of the molecules are assumed to be exactly along the molecular beam axis. In a
practical Stark decelerator, however, the molecules in the beam also have velocity components
perpendicular to the molecular beam axis. The electric fields applied in the decelerator have to
drive the selected molecules back towards the molecular beam axis, resulting in a transverse
oscillatory motion. The influence of the coupling of longitudinal and transverse motion and its
influence on the phase stability have been studied and resonances similar to the parametric
oscillations in AC ion traps~\cite{Alheit:IJMS154:155} and AC traps for neutral
molecules~\cite{Veldhoven:PRL94:083001, Bethlem:PRA74:063403, Schnell:JPCA111:7411,
   Luetzow:PRA77:063402} have been observed~\cite{Meerakker:PRA73:023401}. Using overtones in the
Stark decelerator, the coupling can be changed and improved transmission has been
obtained~\cite{Meerakker:PRA71:053409, Meerakker:PRA73:023401}. In these experiments, only a
fraction of the electric field stages is used for the deceleration and remaining ones can be used
for optimized transverse focusing.

Recently, advanced, more complicated electrode geometries using normal deceleration stages and
quadrupole focusing stages have been proposed in order to allow for the decoupling of transverse
focusing and longitudinal manipulation of the molecular packet \cite{Sawyer:EPJD:4111}, but no
experimental evidence for the improvement due to such electrode geometries has been given yet.
However, the independent manipulation of longitudinal and transverse motion has already been shown
for the deceleration of molecules in high-field-seeking states using the alternating-gradient
decelerator \cite{Bethlem:PRL88:133003, Tarbutt:PRL92:173002, Bethlem:JPB39:R263,
   Wohlfart:PRA77:031404}. The transverse distribution of molecules exiting the decelerator has been
experimentally observed using an imaging system \cite{Bethlem:JPB39:R263}. Here we demonstrate that
an array of electrodes in alternating-gradient geometry can also be used for the focusing and
deceleration of molecules in lfs states.

\begin{figure}
   \centering%
   \includegraphics[width=\figwidth]{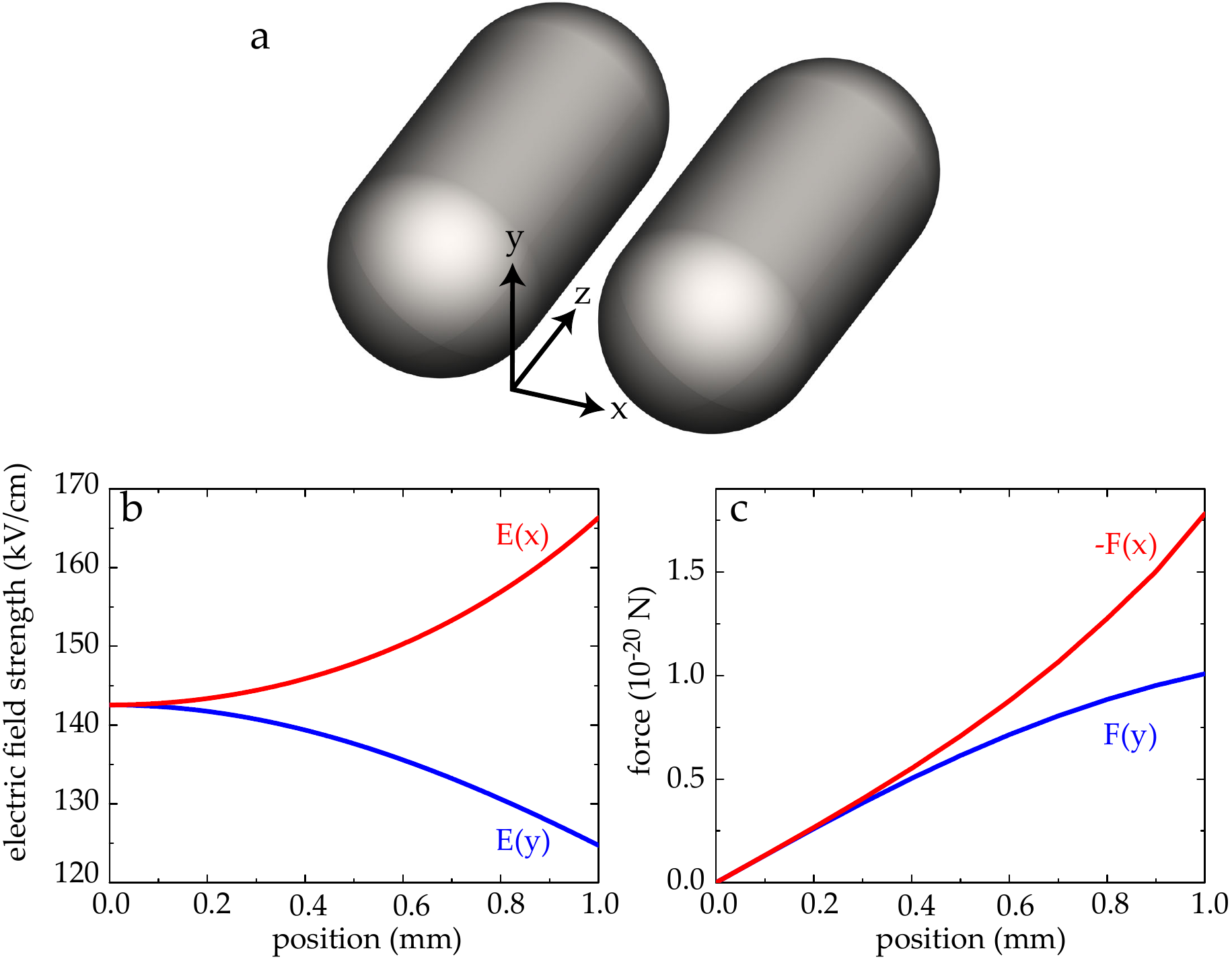}%
   \caption{(Color online) (a) Electrode pair geometry. In every stage two parallel electrodes are
      oriented along the molecular beam axis. (b) The norm of the electric field created by the pair
      of electrodes in the two transverse directions. (c) The resulting forces for OH radical in its
      lfs \lfsstate{} state along the $x$ and $y$ axes. The negative of the force along the $x$ axis
      is plotted to ease the comparison of the strengths of focusing and defocusing forces.}
   \label{fig:geometry}
\end{figure}

We have set up an alternating-gradient Stark decelerator in which the electrodes creating the
electric fields are oriented along the molecular beam axis, with pairs of two parallel electrodes
per stage, as shown in figure~\ref{fig:geometry}\,a.
The decelerator consists of successive pairs of electrodes which are rotated by 90\degree\ around
the molecular beam axis. In this electrode arrangement the molecules are decelerated in the space
between two successive electrode pairs, very similar to the normal Stark decelerator. In addition,
however, this geometry allows to independently provide transverse focusing when the molecular packet
is inside the electrode pair, where the electric field gradient in longitudinal direction vanishes.
One has to take into account that for an individual stage the force is focusing in one transverse
direction but defocusing in the other, as shown by the electric field norm and focusing force given
in figure~\ref{fig:geometry}\,b and c: Molecules in lfs states are focused in the plane of the
electrodes and are defocused in the plane perpendicular to the electrodes. Therefore, electrode
pairs are arranged in perpendicular transverse planes and AG focusing is used to transport molecules
through the beamline, similar to dynamic focusing in charged particle accelerators.

Here, we apply AG deceleration for the deceleration of OH in its lfs \lfsstate{} state. Using a
prototype setup, consisting of only 27 stages, we could remove 50~\% of the kinetic energy from the
molecular packets in a supersonic jet. The independent control over the transverse focusing allows
to adopt the focusing strength of the electric field stages to the velocity of the molecular packet
and, therefore, to keep the transverse focusing and the transverse acceptance constant over the
whole deceleration process. In addition, using exactly the same experimental setup, we have also
decelerated OH radicals in their hfs \hfsstate{} state. This lower $\Lambda$-doubling component of
the rotational ground state is the absolute ground state of OH. These experiments demonstrate the
versatility of the AG decelerator, which can be used for the deceleration of molecules in any polar
quantum state.

By applying different bias voltages to the decelerator electrodes, which provide a minimum electric
field strength to the molecules throughout the whole deceleration process, we could observe the
level crossing of two hyperfine states of OH in its hfs \hfsstate{} state. This level crossing
manifests itself in a loss channel for the decelerated packet, which can be suppressed by applying
bias voltages large enough to always stay above the field strength of the crossing.

\section{Experimental details}
\label{sec:experimental}

\begin{figure}
   \centering
   \includegraphics[width=\figwidth]{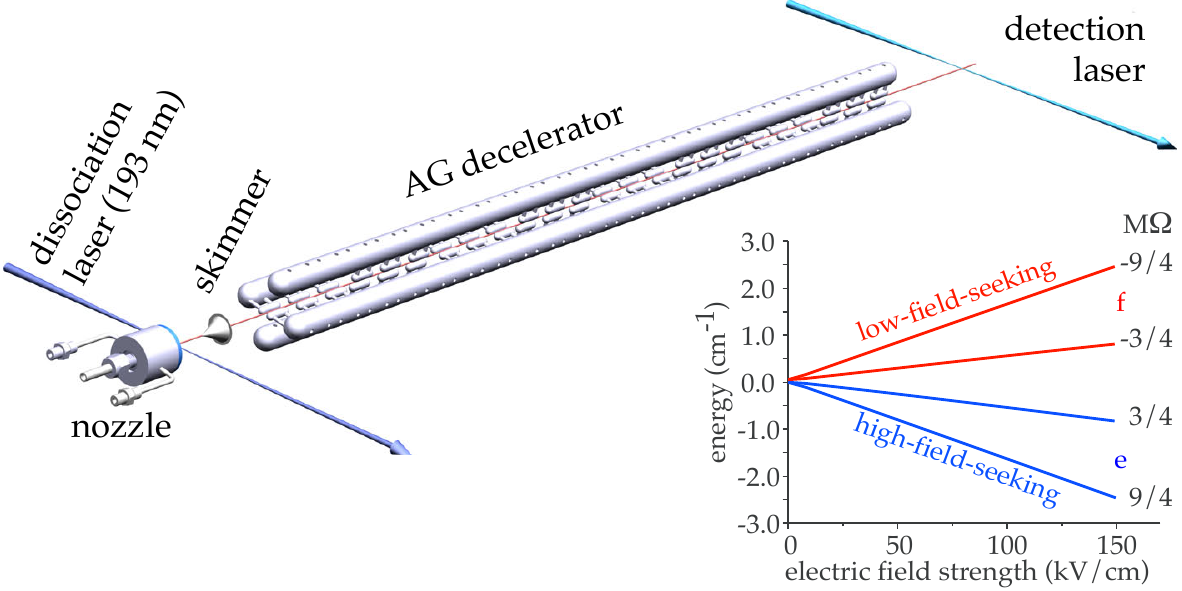}
   \caption{(Color online) Scheme of the experimental setup. In the inset the Stark shifts of the
      different $M\Omega$ manifolds of OH radicals in their rovibronic \state{} are shown. See text
      for details.}
   \label{fig:setup}
\end{figure}

The experiments described here are performed in the same molecular-beam machine that was previously
used for the deceleration of benzonitrile (C$_7$H$_5$N)~\cite{Wohlfart:PRA77:031404}. This setup is
schematically shown in figure~\ref{fig:setup}.
A pulsed beam of OH radicals is produced by photodissociation (193~nm) of \hnooo{} molecules that
are co-expanded with xenon. 20~mm downstream the radicals pass a 1.5~mm diameter skimmer, which is
mounted directly on a gate valve~\cite{Kuepper:RSI77:016106} to allow for a short
nozzle-to-decelerator distance, and enter a second differentially pumped vacuum chamber. The first
electrode pair of the decelerator starts 37~mm behind the tip of the skimmer. The decelerator
consists of 27 electrode pairs arranged along the molecular beam axis. Each electrode has a length
of 13~mm, a diameter of 6~mm, and the spherical end caps of the electrodes have a radius of 3~mm.
The two electrodes of every stage have a distance of 2~mm, successive electrode pairs are placed at
center-to-center distances of 20~mm along the molecular beam axis $z$ and successively rotated by
90\degree{} after every three electrode pairs. The total length of the decelerator is 533~mm. For
the two opposing electrodes of every pair voltages are switched between $\pm15$~kV, corresponding to
a maximum electric field of 142.5~kV/cm on the molecular beam axis, and a bias voltage of typically
$\pm0.3$~kV. The exact field strength values are given in figure~\ref{fig:geometry}\,b. OH radicals
are detected 655~mm behind the laser production by laser-induced fluorescence (LIF) using
time-resolved photon-counting. In the experiments presented here we use a frequency-stabilized
continuous-wave ring-dye laser for the electronic excitation
($A\,{}^2\Sigma^+\leftarrow{}X\,{}^2\Pi_{3/2}$) of the molecules. Its narrow linewidth (1~MHz)
allows to specifically detect molecules in a given hyperfine state. At the same time, we can record
the complete arrival time distribution of each molecular packet, as the laser excitation and the
photon detection are continuous. In that way complete time-of-flight (TOF) profiles of the molecular
packets from the nozzle to the detector can be recorded in a single shot experiment. For the
experimental TOF profiles shown in this paper the data from 8000 experiments are averaged,
corresponding to a measuring time of approximately 7~min at 20~Hz.

\section{Switching schemes}
\label{sec:switching-schemes}

The alternating-gradient decelerator allows to individually choose the transverse focusing and the
deceleration strength, both for molecules in lfs and in hfs states~\cite{Bethlem:JPB39:R263}. To
describe the sequence of times at which the high voltages are switched, we use the concept of a
synchronous molecule, which is by definition always at the same position within an AG lens when the
high voltages are switched. For the deceleration of molecules in lfs states, one must let the
molecular packet fly from regions of low electric field into regions of high electric fields. In
order to achieve phase stability the field must be switched off on the rising flank of the field,
\ie, before the molecular packet enters the region of constant electric field along the molecular
beam axis inside the pair of electrodes. That way, faster molecules are decelerated more and slower
ones less than the synchronous molecule, resulting in an oscillation of the molecules in the packet
around the synchronous molecule, and the packet is effectively kept
together~\cite{Bethlem:PRA65:053416}. Here, we have applied a switching pattern based on the ones
used in the normal Stark decelerator: when the synchronous molecule is 2~mm before the position of
minimum electric field (the center between two successive electrode pairs), the fields on both pairs
of electrodes are switched on. When the field is again switched off 2~mm behind the center position,
the molecular packet is bunched longitudinally~\cite{Crompvoets:PRL89:093004}. If the field is kept
on longer, however, it is decelerated, as the packet now has to climb more of the potential hill
into the electrode pair. To achieve transverse focusing the fields are then switched on and off once
more for every stage when the molecules are inside the electrode pair.

For the first high-voltage pulse in every stage, the position along the molecular beam axis $z$
where we switch off the electric field for the synchronous molecule is called $d$; this position is
specified relative to the minimum of the electric field in between two neighboring electrode pairs.
The length over which the field is applied is $b$. These values $d$ and $b$ determine the amount of
deceleration and longitudinal focusing (bunching), respectively. In principle, $b$ determines also
the length of the molecular packet that is bunched, but this effect is negligible here due to the
small length of the initial packet. The second high-voltage pulse is applied symmetrically around
the center of the electrode pair and the fields are switched on over a distance $f$ for the
synchronous molecule. The length of this pulse determines the amount of transverse focusing. These
parameters are graphically depicted next to the experimental results given in
section~\ref{sec:results}.

For maximum longitudinal acceptance the amount of energy removed from the synchronous molecule per
stage should be constant and, therefore, $d$ should be constant over the whole decelerator.
Accordingly, the transverse focusing should also be constant over the decelerator and, therefore,
the molecules should spend the same time in the focusing field for every stage. In previous AG
deceleration experiments~\cite{Bethlem:PRL88:133003, Tarbutt:PRL92:173002, Bethlem:JPB39:R263,
   Wohlfart:PRA77:031404} the velocity change was small and a constant focusing length $f$ could be
used. Since we achieve a larger velocity change in the experiments presented here, the focusing
parameter $f$ has to be lowered over the course of the deceleration to compensate for the velocity
changes. Generally, a constant amount of energy is removed from the molecular packet per
deceleration stage, resulting in a quadratic reduction of velocity during the deceleration process.
In order to obtain constant transverse focusing, which provides the maximum transverse acceptance,
$f$ should also be reduced quadratically during the deceleration process. As a first order
approximation, we have experimentally optimized a linearly changing $f$, resulting in an increase of
the peak intensities of the decelerated molecular packets of up to 20~\% relative to the intensities
obtained using switching sequences with a constant $f$. However, when the velocity changes during
the deceleration process will be even larger than demonstrated here, quadratic changes of $f$ will
become crucial. Additionally, it has to be taken into account that the deceleration pulse has also a
transverse focusing effect. This effect is larger if we let the molecular packet fly further towards
the electrodes, \ie, for stronger deceleration. Therefore, the amount of focusing $f$ has generally
to be lowered for larger $d$. The actual parameter values used in the experiments are depicted
alongside the experimental data described in section~\ref{sec:results}.

For the deceleration of packets of OH radicals in their hfs \hfsstate{} state, switching sequences
as described before are used \cite{Wohlfart:PRA77:031404}. We characterize the switching sequences
by a $d$ and $f$ parameter. For the synchronous molecule, $d$ is the position along the molecular
beam axis $z$ where the electric field is switched off in every electric field lens, relative to the
center of the electrode pairs, with positive values running towards the detector. $f$ is the length
over which the electric field is switched on. It determines the amount of transverse focusing, where
a larger value corresponds to stronger focusing, and the $d$ parameter describes amount of
deceleration, with larger values corresponding to more deceleration.

\section{Trajectory calculations}
\label{sec:trajectory-calc}

\begin{figure}
   \centering
   \includegraphics[width=\figwidth]{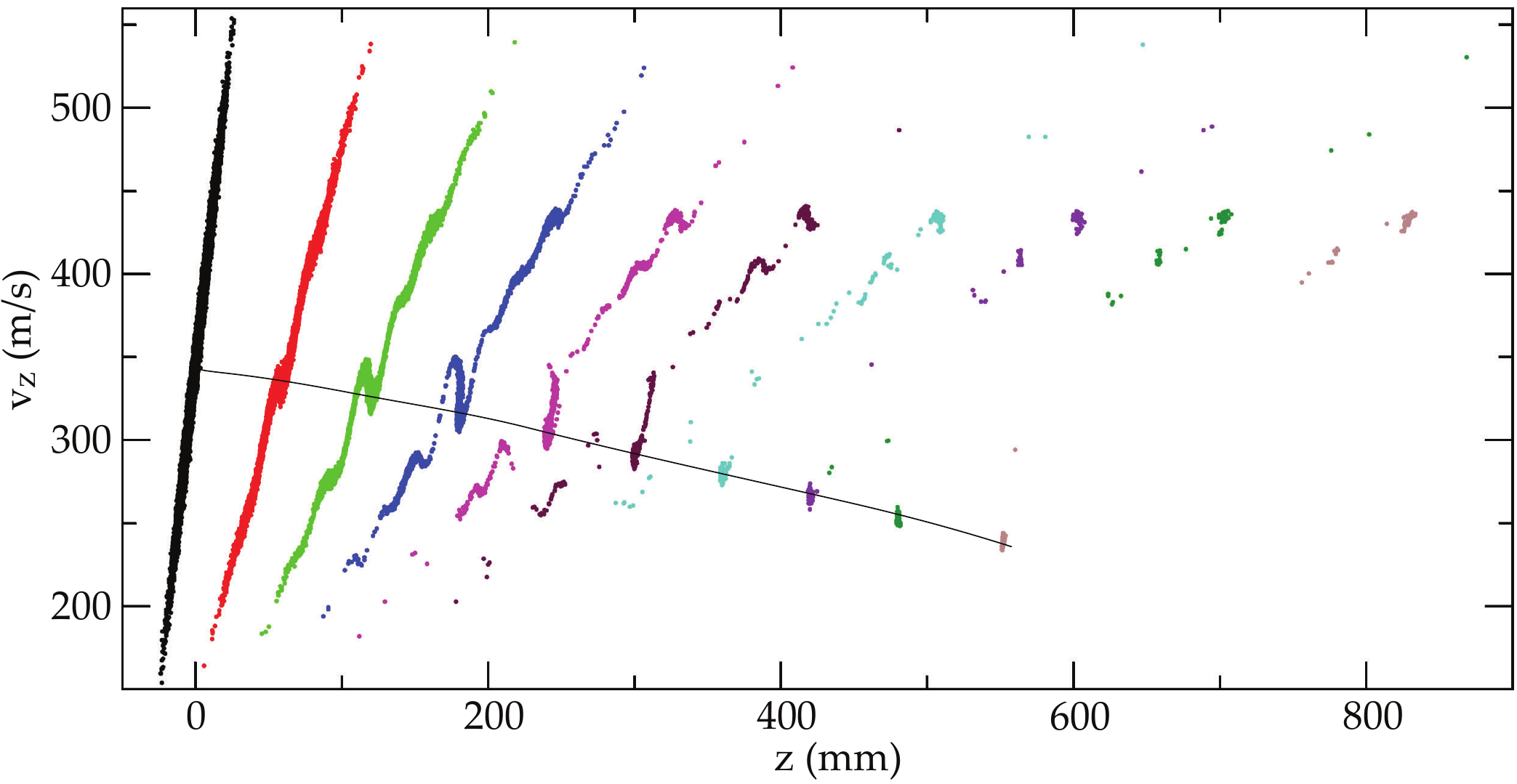}
   \caption{(Color online) Phase-space evolution during the deceleration of OH radicals in the
      low-field-seeking \ensuremath{\lfsstate,M\Omega=-9/4} state from 345~m/s to 239~m/s. The
      longitudinal phase-space distributions are shown before the decelerator and at the center of
      field lenses $3,6,9\ldots$ To guide the eye, the slowing of the accepted packet is indicated
      by the black line.}
   \label{fig:phasespace:evolution}
\end{figure}
\begin{figure}
   \centering%
   \includegraphics[width=\figwidth]{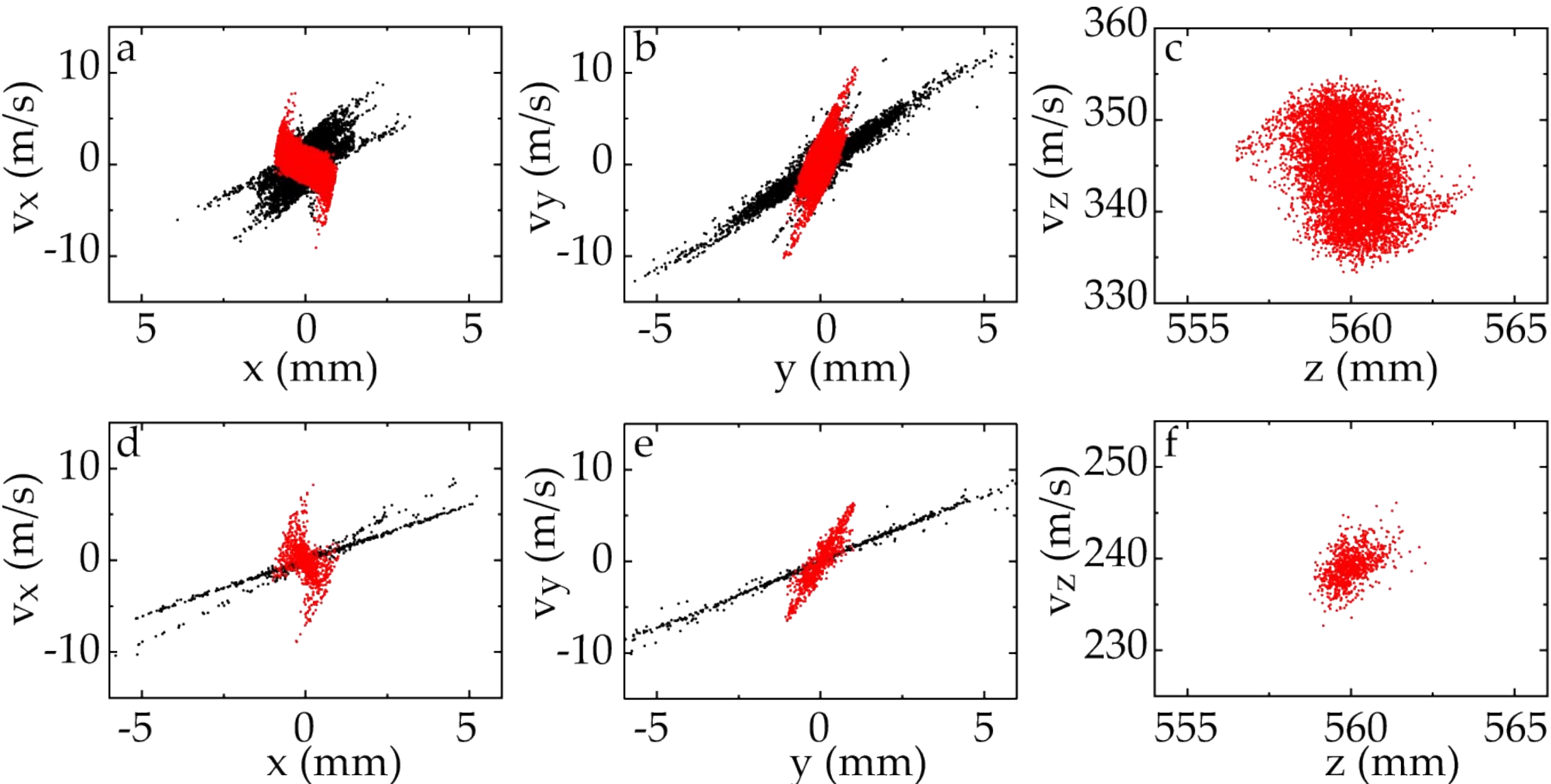}
   \caption{(Color online) Simulated phase-space distributions for the low-field-seeking
      \ensuremath{\lfsstate,M\Omega=-9/4} state of OH 20~mm behind the center of the last AG lens.
      The upper row shows the two transverse and the longitudinal phase-space distributions for
      focusing at 345~m/s and the lower row shows the distributions for deceleration from 345~m/s to
      239~m/s. The gray (red) phase-space areas depict the molecules in the captured packet. For the
      transverse distributions all other molecules -- which are not captured, but pass the
      decelerator on metastable trajectories -- are shown in black.}
   \label{fig:phasespace-lfs}
\end{figure}
\begin{figure}
   \centering%
   \includegraphics[width=\figwidth]{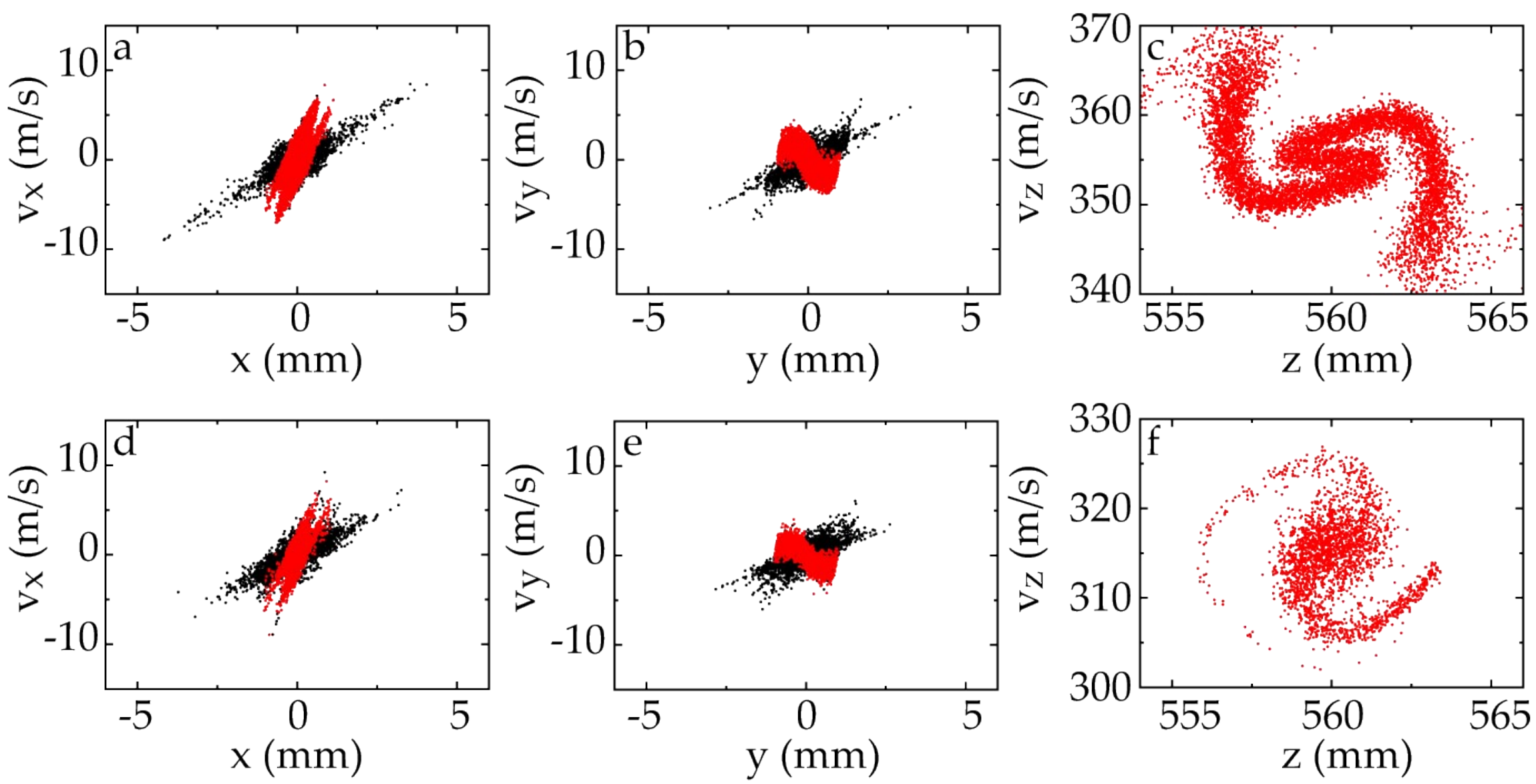}
   \caption{(Color online) Simulated phase-space distributions for the high-field-seeking
      \ensuremath{\hfsstate,M\Omega=+9/4} state of OH 20~mm behind the center of the last AG lens.
      The upper row shows the two transverse and the longitudinal phase-space distributions for
      focusing at 355~m/s and the lower row shows the distributions for deceleration from 355~m/s to
      316~m/s. The gray (red) phase-space areas depict the molecules in the captured packet. For the
      transverse distributions all other molecules -- which are not captured, but pass the
      decelerator on metastable trajectories -- are shown in black.}
   \label{fig:phasespace-hfs}
\end{figure}

Corresponding trajectory calculations to simulate the TOF profiles are performed using the
libcoldmol program package \cite{Kuepper:libcoldmol}. In the simulations, we have included all
hyperfine levels of the respective $\Lambda$-doubling component of OH, \ie, the $|M\Omega|=9/4$ and
the less polar $|M\Omega|=3/4$ states (see figure~\ref{fig:setup}), using the appropriate weighting
due to the number of hyperfine states belonging to these two manifolds. These simulations also yield
the phase-space distributions of the molecules at any time during the deceleration process. In
figure~\ref{fig:phasespace:evolution} the phase-space distributions for the times when the
synchronous molecule is at the center of every third electrode are plotted, starting 20~mm before
the center of the first electrode. Transverse and longitudinal phase-space distributions of OH
radicals in the lfs \ensuremath{\lfsstate,M\Omega=-9/4} state and the hfs
\ensuremath{\hfsstate,M\Omega=+9/4} state at the detector are shown in
figures~\ref{fig:phasespace-lfs} and \ref{fig:phasespace-hfs}, respectively. The distributions of
molecules in the captured packets are shown in gray (red). Due to the finite length of the
decelerator molecules can also reach the detector on metastable trajectories. For the transverse
distributions these molecules are shown in black.

The phase-space acceptance of the decelerator for a given switching sequence can be derived from
trajectory calculations.
\begin{figure}
   \centering%
   \includegraphics[width=\figwidth]{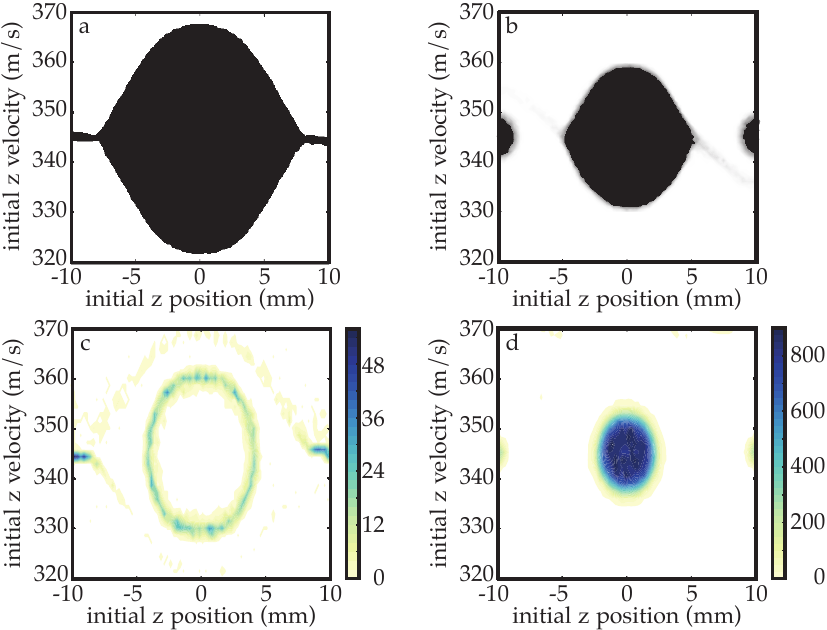}
   \caption{(Color online) The longitudinal phase-space acceptance for focusing OH radicals in their
      low-field seeking $\lfsstate,M\Omega=-9/4$ state. a and b are the acceptance for a
      one-dimensional model along the molecular beam axis, c and d display the results of the full
      three-dimensional calculations, where the density (molecules/(mm$\cdot$m/s)) of the projection
      of the phase-space acceptance onto the $(z,v_z)$ plane is displayed. In a and c high-voltages
      are switched analogous to a normal Stark decelerator experiment, whereas in b and d the
      simulations are performed with the same switching sequence for guiding at 345~m/s as used in
      the experiments of this work. See text for details.}
   \label{fig:acceptance}
\end{figure}
In figure~\ref{fig:acceptance} the longitudinal phase-space acceptances of the AG decelerator for OH
radicals in their low-field seeking $\lfsstate,M\Omega=-9/4$ state are shown for different
high-voltage sequences applied. All phase-space acceptances are calculated by numerical trajectory
calculations using the correct electric fields and field gradients obtained using finite element
methods (Comsol Multiphysics). For each set of parameters $3\cdot10^{8}$ trajectories from a uniform
distribution of an initial six-dimensional phase-space area of $1.8\cdot10^6$~(mm$\cdot$m/s)$^3$
through a 2~m long decelerator are calculated. In all panels the acceptance along the molecular beam
axis $z$ is plotted. The upper two images (a and b) show the phase-space acceptance obtained from
calculations where all molecules are confined on the molecular beam axis. Generally, in such a
calculation all molecules eventually reach the detector. In figure~\ref{fig:acceptance}\,a only the
initial phase-space positions of molecules that reach the detector within the synchronous packet,
the packet around the synchronous molecule, are plotted. In figure~\ref{fig:acceptance}\,b the
initial phase-space positions of molecules that reach the detector within the synchronous packet and
of molecules in the packets half an electrode pair ahead and behind are plotted (\emph{vide infra}).
In figure~\ref{fig:acceptance}\,c and d the corresponding densities of the projections onto the
$(z,v_z)$ plane of initial phase-space positions of the molecules reaching the detector in full
three-dimensional simulations are plotted.

When the AG decelerator would be used analogous to a normal Stark decelerator for the deceleration
of OH in its lfs $\lfsstate,M\Omega=-9/4$ state, the high voltages are only switched on when the
synchronous molecule is between successive electrode pairs. This results in the one- and
three-dimensional phase-space acceptances shown in figure~\ref{fig:acceptance}\,a and c,
respectively. In the three-dimensional calculation, only a small ring of accepted initial
phase-space positions is observed. For most phase-space positions, \ie, the central part of the
one-dimensional distribution, the process is not stable. The observed halo and the central region of
instability have been described before for the normal Stark decelerator and the halo has been
ascribed to instabilities due to the coupling of longitudinal and transverse
motion~\cite{Meerakker:PRA73:023401}. In the normal Stark decelerator this problem can be mitigated
by applying overtone switching frequencies.

However, the AG decelerator allows to independently control the deceleration and the transverse
focusing strength, allowing to minimize losses due to such parametric amplification from the
coupling of longitudinal and transverse motion. The result of applying an additional transverse
focusing field around the center of the AG lenses is shown in figure~\ref{fig:acceptance}\,b and d:
Here, the acceptance is calculated using the switching sequence applied in the experiments described
below. In the one-dimensional calculation the packet is split up into two somewhat smaller packets
due to the smaller lattice-cell lengths: The molecules feel the force of the electric fields every
10~mm instead of every 20~mm due to the doubled number of high-voltage pulses -- half of them for
longitudinal bunching, the other half for transverse focusing. Since ``the molecules do not know
this'', they might use the fields intended for bunching for focusing instead, and vice versa. For
the three-dimensional calculation under these conditions the projection of the initial phase-space
density onto the $(z,v_z)$ plane is highest around the synchronous molecule, allowing to efficiently
couple a molecular beam, which typically has a Gaussian-like phase-space distribution, into the AG
decelerator.

From these trajectory calculations the phase-space acceptance for guiding OH radicals in their lfs
state at 345~m/s is determined to $10^{3}$~(mm$\cdot$m/s)$^3$. This is one order of magnitude
smaller than the corresponding phase-space acceptance of a normal Stark decelerator with similar
dimensions, \ie, 2~mm electrode spacing and 5.5~mm lens-to-lens distance, and the same voltages
applied.

\section{Experimental results}
\label{sec:results}

\begin{figure}
   \centering
   \includegraphics[width=\figwidth]{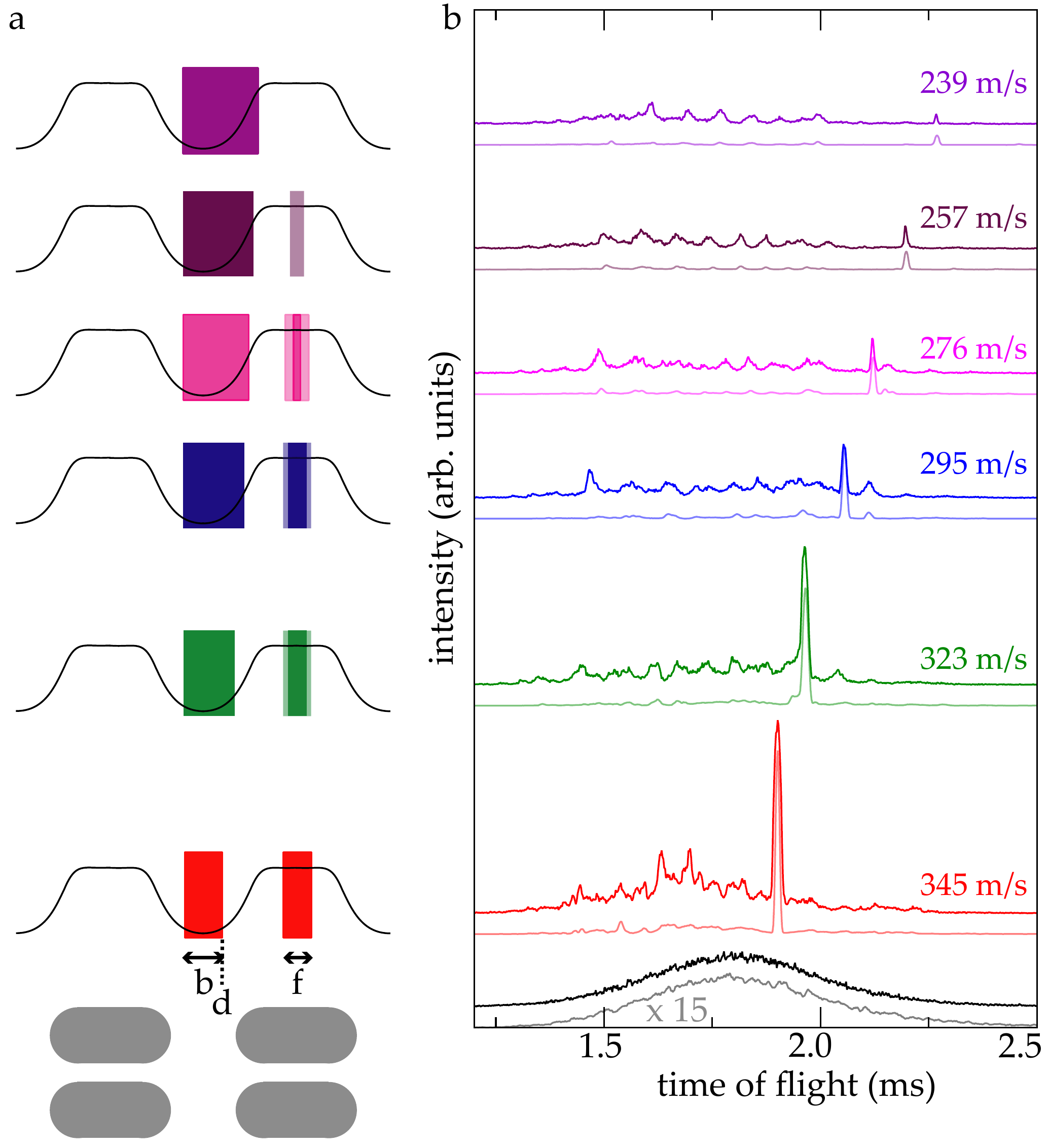}
   \caption{(Color online) a) Graphical representations of the high-voltage switching parameters
      $b$, $d$, and $f$, see text for details. b) Deceleration sequence of OH in its low-field
      seeking \lfsstate{} state, with the simulated time-of-flight profiles underneath the
      experimental measurements for different focusing and deceleration parameters. The lowest
      traces are free-flight profiles. The second (red) traces show the arrival-time distribution
      for focusing at a constant velocity of 345~m/s. The following traces are TOFs for deceleration
      from 345~m/s to successively lower final velocities, as noted in the graph.}
   \label{fig:deceleration:lfs}
\end{figure}

The experimental results for the deceleration of OH in its lfs \lfsstate{} state are shown in
figure~\ref{fig:deceleration:lfs}, together with their respective simulations.
Comparing the TOF distributions for free-flight and guiding measurements, shown in the lowest two
traces of figure~\ref{fig:deceleration:lfs}\,b, the improved peak intensity of the focused packet
relative to the unfocused packet is obvious. Moreover, due to the velocity selectivity of the
focusing the velocity distribution of the accepted packet is considerably narrower, corresponding to
a colder packet of molecules. This can also be seen in the phase-space distributions shown in
figure~\ref{fig:phasespace-lfs}. In all deceleration measurements we have applied a linear variation
of $f$ over the decelerator, which resulted in a maximum improvement of 21~\% compared to a constant
$f$ for the strongest deceleration sequences. The variation of $f$ is depicted in
figure~\ref{fig:deceleration:lfs}\,a by the shaded areas. For the strongest deceleration sequence we
could decelerate OH from 345~m/s to 239~m/s, removing more than 50~\% of the kinetic energy. In
additional experiments (not shown) we have also decelerated packets of OH in its lfs state from
305~m/s to 199~m/s~\cite{Wohlfart:thesis:2008}, the slowest velocity obtained for molecular packets
from an AG decelerator so far.

\begin{figure}
   \centering
   \includegraphics[width=\figwidth]{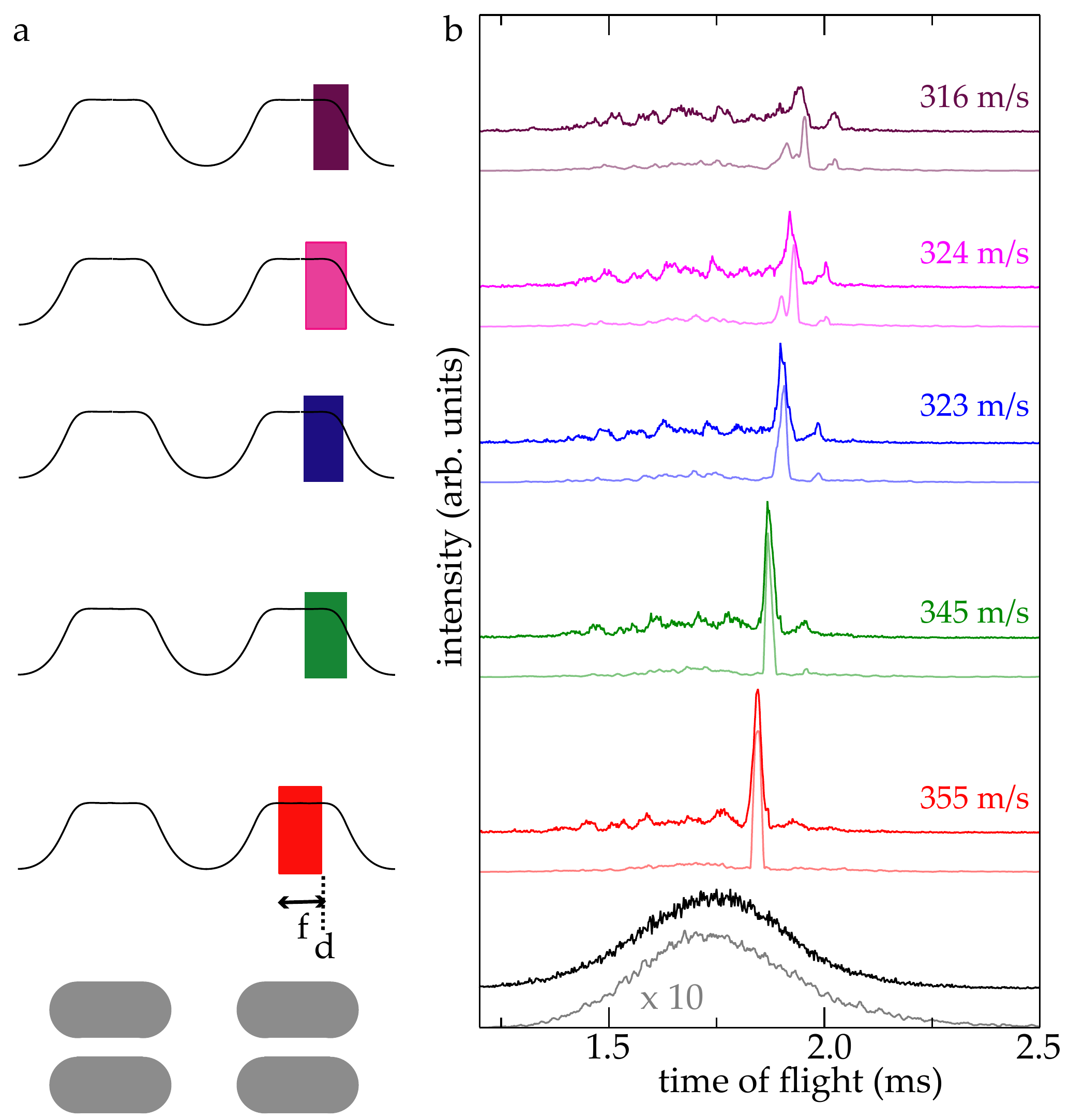}
   \caption{(Color online) a) Graphical representations of the high-voltage switching parameters $d$
      and $f$. b) Deceleration sequence of OH in its high-field seeking \hfsstate{} state, with the
      simulated time-of-flight profiles underneath the experimental measurements for different
      focusing and deceleration parameters. The lowest traces are free-flight profiles. The second
      (red) traces show the arrival-time distribution for focusing at a constant velocity of
      355~m/s. The following traces are TOFs for deceleration from 355~m/s to successively slower
      final velocities, as noted in the graph.}
   \label{fig:deceleration:hfs}
\end{figure}

The TOF profiles for the deceleration of OH in its hfs \hfsstate{} state are shown in
figure~\ref{fig:deceleration:hfs}.
When letting the molecular packet fly out of the electrode pair while the electrodes are powered, an
overall reduction of the molecular velocity results in a delayed arrival at the detector. For
increased $d$, the final velocity is decreased further, down to a final velocity of 316~m/s for
$d=5.25$~mm. For this strongest deceleration the kinetic energy of the molecules is reduced by
approximately 21~\%.

For all simulations, the intensity ratios of the measurements with an electric field applied match
the experimental results rather well. They do, however, predict a considerably larger intensity of
the focused beams compared to the free flight measurements than is observed experimentally. We
ascribed this to mechanical misalignment of the high-voltage electrodes. When we assume the
electrode pairs to be shifted in a random fashion in all three dimensions, using a Gaussian
distribution with a standard deviation of $\pm150$~$\mu$m, a decrease of the simulated intensities
is obtained that matches the experimental results.

\begin{figure}
   \centering
   \includegraphics[width=\figwidth]{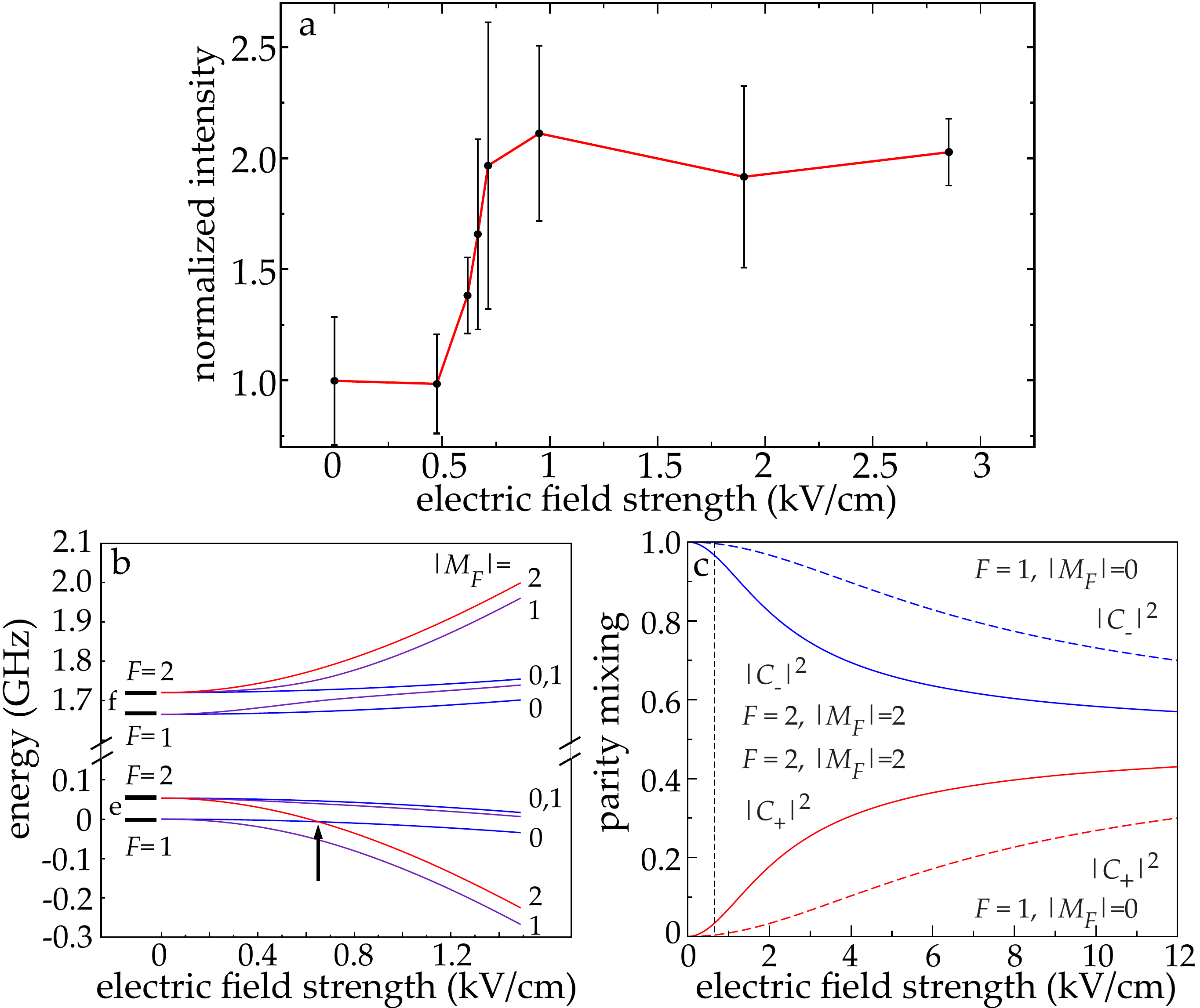}
   \caption{(Color online): (a) Bias-voltage dependence of the transmission for OH in its hfs
      \hfsstate{}, $F=2$ state. In all measurements the same switching sequence is applied and the
      integrated intensity of the focused molecular packet is plotted against the nominal electric
      field strength inside the electric field stages. (b) Stark energies of the individual
      hyperfine states of OH in its \state{} rotational state. (c) Parity components of the two
      hyperfine states that cross at 640~V/cm. See text for details.}
   \label{fig:bias}
\end{figure}

In addition, molecules can be lost from the polar quantum states due to Majorana transitions or
diabatic following of potential energy curves. These effects are minimized by applying $\pm300$~V
bias voltage to the electrodes instead of switching to ground, as described above and used in all
experiments shown here. For the experiments on OH in its lfs states, a steady signal increase up to
a factor of 3.5 can be observed when tuning the bias voltage from 0~V to $\pm 300$~V. We ascribe
this to transverse focusing effects of the bias field. For the hfs states the integrated intensities
of the synchronous packet is given as a function of the bias field strength in
figure~\ref{fig:bias}\,a. For a threshold voltage of 67~V, which corresponds to an electric field
strength at the center of the AG lenses of 640~V/cm, a sudden rise in signal intensity is observed.
At this field strength a real crossing between two hyperfine states occurs, which is marked by the
arrow in figure~\ref{fig:bias}\,b. One of these levels belongs to the $F=2,M\Omega=9/4$ manifolds,
whereas the other level belongs to the $F=1,M\Omega=3/4$ manifolds. Both hyperfine states have
negative parity. The coupling of states with the same parity is dipole forbidden. In an electric
field, however, hyperfine states of different parities mix. In figure~\ref{fig:bias}\,c, the squares
of the mixing coefficients for both hyperfine levels of the crossing are shown. For positive parity,
the mixing ratios at an electric field strength of 640~V/cm are 1~\% and 4~\% for the
\ensuremath{\hfsstate,F=1,M_F = 0} and \ensuremath{\hfsstate,F=2,M_F = 2} state, respectively. This
parity mixing for bias voltages below the threshold can be sufficient to allow transitions from the
\ensuremath{F=2,M_F = 2} state to the \ensuremath{F=1,M_F = 0} state. In our experiment, only the
$F=2,M_F = 2$ state is detected, and, therefore, the detected intensity is reduced. However, it is
obvious that the losses can be minimized by applying a bias voltage just above the field strength at
which the crossing occurs.

\section{Conclusions}
\label{sec:conclusions}

The alternating-gradient focusing and deceleration of OH in both high-field-seeking and
low-field-seeking states of its rovibronic ground state have been demonstrated using a single
experimental setup. This work demonstrates the versatility of the AG decelerator, which can, in
principle, be used for the deceleration of molecules in any polar quantum state.

For OH radicals in the lfs \lfsstate{} state more than 50~\% of the kinetic energy has been removed
in deceleration experiments using an AG decelerator with 27 stages. Moreover, it has been
demonstrated that the AG decelerator allows to separately change the transverse and longitudinal
focusing properties. The deceleration achieved in this work has to be compared to the deceleration
of OH using a normal Stark decelerator, where OH in the lfs state has been decelerated to a
standstill, using 108 deceleration stages, and finally trapped using an electrostatic
trap~\cite{Meerakker:PRL94:023004}. Due to the small number of electrodes of the AG decelerator,
deceleration to a standstill was not possible. Moreover, the phase-space acceptance of the AG
decelerator is an order of magnitude smaller than for the normal Stark decelerator.

For the deceleration of the OH radicals in their hfs absolute ground state (\hfsstate), the same
switching sequences as used in previous experiments have been applied~\cite{Bethlem:PRL88:133003,
   Tarbutt:PRL92:173002, Bethlem:JPB39:R263, Wohlfart:PRA77:031404} and about 21~\% of the
molecules' kinetic energy has been removed. In addition, the dependence of the transmission through
the decelerator on an applied bias voltage has been studied, and for OH in its hfs \hfsstate{} state
a threshold behavior has been found. This is attributed to a real crossing of hyperfine states at
the corresponding electric field strength.

\begin{acknowledgments}
   We acknowledge helpful discussions with Sebastiaan Y.\ T.\ van de Meerakker and Hendrick L.\
   Bethlem, as well as financial support from the \emph{Deutsche Forschungsgemeinschaft} within the
   priority program 1116 ``Interactions in ultracold gases''.
\end{acknowledgments}

\bibliographystyle{jk-apsrev}
\bibliography{string,mp}

\end{document}